\documentclass[twocolumn,showpacs,amsmath,amssymb]{revtex4}

\usepackage{graphicx}
\usepackage{dcolumn}
\usepackage{bm}

\begin{document}

\title{
Fault-tolerant Quantum Computation
with Highly Verified Logical Cluster States
}

\author{Keisuke Fujii}
\author{Katsuji Yamamoto}
\affiliation{
Department of Nuclear Engineering, Kyoto University, Kyoto 606-8501, Japan}

\date{\today}

\begin{abstract}
We investigate a scheme of fault-tolerant quantum computation
based on the cluster model.
Logical qubits are encoded by a suitable code
such as the Steane's 7-qubit code.
Cluster states of logical qubits are prepared by post-selection
through verification at high fidelity level,
where the unsuccessful ones are discarded without recovery operation.
Then, gate operations are implemented
by transversal measurements on the prepared logical cluster states.
The noise threshold is improved significantly
by making the high fidelity preparation and transversal measurement.
It is estimated to be about $ 3 \% $ by a numerical simulation.
\end{abstract}


\maketitle

In order to implement reliable computation in physical systems,
either classical or quantum,
the problem of noise should be overcome.
Then, fault-tolerant quantum computation
with error correction has been investigated
in the literature
\cite{Shor95,CaldShor96,Shor96,Stea96,DiViShor96,Stea97,Stea98}.
In the usual quantum error correction (QEC), error syndromes are detected
on encoded qubits, and the errors are corrected according to them.
The noise thresholds for fault-tolerant computation
based on the circuit model are calculated
to be about $10^{-6} - 10^{-3}$
depending on the QEC protocols and noise models
\cite{Stea97,Stea98,Stea99,Stea03,Kitaev97,Preskill98,Knill98,
Gottesman97,Gottesman98,AB-O99}.
A main motivation for QEC comes from the fact that
in the circuit model we must continue to use the original qubits
through computation even if errors occur on them.

On the other hand, more robust computation may be performed
in measurement-based quantum computers
\cite{GC99,ZLC00,OWC,Niel03,Knill05a,Knill05b}.
Teleportation from old qubits to fresh ones
is made by measurements implementing gate operations,
and the original qubits are not retained in this sort of computers.
Then, even recovery operation may not be required.
An interesting computation model
with error-correcting teleportation is proposed
based on encoded Bell pair preparation and Bell measurement,
which provides a high noise threshold $ \sim 3 \% $
\cite{Knill05a,Knill05b}.
The cluster model or one-way computer \cite{OWC}
should also be considered for fault-tolerant computation.
A highly entangled state, called a cluster state, is prepared,
and then gate operations are implemented
by measuring the qubits in the cluster
with feedforward for the post-selection of measurement bases.
This gate operation in the cluster model
may be viewed as the one-bit teleportation \cite{ZLC00}.
A promising scheme for linear optical quantum computation is proposed,
where the deterministic gates are implemented
by means of the cluster model \cite{Niel04}.
Fault-tolerant computation is built up for this optical quantum computer
by using a clusterized version of the syndrome extraction for QEC
in the circuit model \cite{Stea97}.
The noise thresholds are then estimated
to be about $ 10^{-3} $ for photon loss
and $ 10^{-4} $ for depolarization \cite{Niel06}.
The threshold result is also argued
by simulating fault-tolerant QEC circuits with clusters
\cite{Rausen03,ND05,AL06}.
These approaches, however, would not provide
the proper threshold in the cluster-model computation.

In this Letter, we investigate a novel scheme
of fault-tolerant quantum computation
by making a better use of the unique feature of quantum processing
in the cluster model,
that is, once clusters are prepared
they are just consumed by measurements for computation.
(In this respect, we may share the concept
with the error-correcting teleportation method
\cite{Knill05a,Knill05b}.)
Specifically, fault-tolerant computation is performed as follows:
\begin{list}{}{}
\item[(I)]
{\it Preparation through verification}.
Clusters of logical qubits are prepared by post-selection
through verification processes
to guarantee high enough fidelity against errors.
\item[(II)]
{\it Transversal measurement}.
Gate operations in the cluster model are implemented
by transversal measurements of logical Pauli operators
on the qubits encoded by a suitable code.
\end{list}
In the preparation process (I)
the error syndrome is detected only for verification,
as described in detail later.
Some number of clusters may be created and verified in parallel,
and the unsuccessful ones are discarded without recovery operation,
which is quite distinct from the usual QEC methods.
This enables the preparation of clean enough clusters by post-selection
to achieve a high noise threshold $ \sim 1 \% $.
The transversal measurement (II) is really performed
on a specific class of stabilizer codes including the Steane's 7-qubit code
\cite{CaldShor96,Stea96,Gottesman98}.
Some relevant Clifford gates, $ H $, $ S $ and C-$ Z $,
operate transversally on such a quantum code
without spreading errors among physical qubits.
The logical measurement in the basis
$ \{ | 0_{L} \rangle , | 1_{L} \rangle \} $
is implemented by the bitwise measurements on physical qubits,
and the measurement basis may be adjusted suitably
with the transversal $ H $ or $ SH $ rotation.
These transversal operations are also used in the preparation process (I).
It will be shown further that some non-Clliford gate,
e.g., the $ \pi / 8 $ gate, for universal computation
can even be implemented by preparing a specific qubit
and making a transversal measurement.

\begin{figure}
\centering
\scalebox{.3}{\includegraphics*[0cm,0cm][20cm,12.5cm]{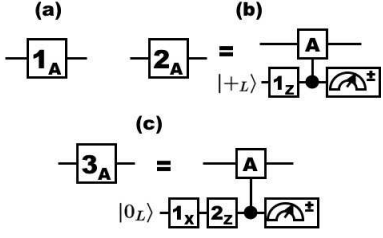}}
\caption{(a) Box 1 of $ A = Z $ ($ X $) measures
the $ Z $ ($ X $) part of the stabilizer.
(b) Box 2 of $ A = X $ ($ Z $) extracts the $ Z $ ($ X $) errors,
and implements the logical measurement of $ A $.
(c) Box 3 of $ A = X $ ($ Z $) extracts the $ Z $ ($ X $) errors
at a higher fidelity, acting ideally as the identical operator.
Transversal measurements on logical qubits are made
in the basis $ \{ | +_{L} \rangle , | -_{L} \rangle \} $.
}
\label{verification}
\end{figure}

We first describe the preparation process (I).
Given significant error rates $ \sim 1 \% $,
we prepare logical cluster states of high fidelity
by post-selection through elaborate verification.
Tools for verification are given in Fig. \ref{verification}
according to fidelity levels.
As the level-1 verification,
the Box 1 of $ A = Z $ ($ X $) in Fig. \ref{verification} (a)
measures the $ Z $ ($ X $) part of the stabilizer
to detect $ X $ ($ Z $) errors
\cite{Shor96,DiViShor96,Stea97,Stea98};
specifically $ | +_{L} \rangle $ ($ | 0_{L} \rangle $)
is verified against $ X $ ($ Z $) errors.
The Box 2 of $ A = X $ ($ Z $) in Fig. \ref{verification} (b)
for the level-2 verification extracts the $ Z $ ($ X $) error syndrome
\cite{Stea97,Stea98,Stea99}.
The combination of C-$ A $ gate, level-1 ancilla $ | +_{L} \rangle $
and $ | \pm_{L} \rangle $ measurement in the Box 2
also implements the logical measurement of $ A $.
(Henceforth the transversal measurements on logical qubits are made
in the basis $ \{ | +_{L} \rangle , | -_{L} \rangle \}
= H \{ | 0_{L} \rangle , | 1_{L} \rangle \} $.)
Then, for the target $ | +_{L} \rangle $ ($ | 0_{L} \rangle $),
which is the + 1 eigenstate of $ X $ ($ Z $),
the outcome of the measurement on the control $ | +_{L} \rangle $
in the Box 2 of $ A = X $ ($ Z $) should be + 1 in the absence of errors.
Hence, by making the error syndrome extraction
and the logical measurement of $ A = X $ ($ Z $) with the Box 2
we can check efficiently the $ Z $ ($ X $) errors
in the level-1 $ | +_{L} \rangle $ ($ | 0_{L} \rangle $)
which is already verified against the $ X $ ($ Z $) errors with the Box 1.
The Box 3 of $ A = X $ ($ Z $) in Fig. \ref{verification} (c)
for the level-3 verification extracts
the $ Z $ ($ X $) error syndrome at a higher fidelity
with a level-2 verified ancilla $ | 0_{L} \rangle $.
It acts ideally as the identical operator.

\begin{figure}
\centering
\scalebox{.3}{\includegraphics*[0cm,0.5cm][20cm,7.5cm]{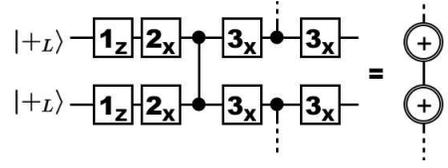}}
\caption{Fault-tolerant preparation of cluster states.
Level-2 $ | +_{L} \rangle $'s verified by Box 1$ {}_Z $ and Box 2$ {}_X $
are connected with C-$ Z $ gates,
and verified further by Box 3$ {}_X $'s.}
\label{cluster-pre}
\end{figure}

By using these verification tools,
we prepare fault-tolerantly logical cluster states of high fidelity,
as shown in Fig. \ref{cluster-pre}.
Logical qubits $ | +_{L} \rangle $ and $ | 0_{L} \rangle $
are encoded with physical qubits as given in Refs. \cite{Stea97,Stea98}.
They are subsequently verified against $ X $ and $ Z $ errors
by Box 1 and Box 2,
as seen in Fig. \ref{verification} (c) and Fig. \ref{cluster-pre}.
Then, level-2 $ | +_{L} \rangle $'s
are connected with transversal C-$ Z $ gates
to construct the required clusters.
The Box 3$ {}_X $'s are placed after each C-$ Z $ connection
for the level-3 verification.
The $ Z $ errors in the qubit $ | +_{L} \rangle $
is detected by the one Box 3$ {}_X $,
while the $ X $ errors, which propagate as $ Z $ errors
through the C-$ Z $ gate, are detected by the other Box 3$ {}_X $.
If any error is detected in these verification processes,
the unsuccessful clusters are discarded.
Some number of clusters may be created and verified in parallel
so that the required clean clusters are obtained
sufficiently by post-selection.

\begin{figure}
\centering
\scalebox{.3}{\includegraphics*[0cm,0cm][28cm,7cm]{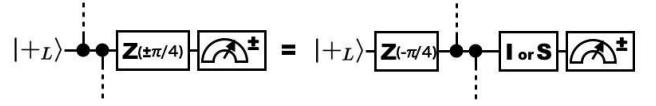}}
\caption{Implementation of the $ \pi / 8 $ gate
by preparation of $ Z (- \pi / 4 ) | + \rangle $
and transversal measurement.}
\label{pi8-ope}
\end{figure}

We next consider the transversal measurement (II)
for universal computation.
In the cluster model, the operation
$ H Z( \theta ) = H e^{-i \theta Z / 2} $
is implemented by the measurement in the basis
$ Z( \pm \theta ) \{ | + \rangle , | - \rangle \} $
to be post-selected by the outcome of preceding measurements
\cite{OWC}.
Non-Clifford gates, e.g., the $ \pi / 8 $ gate = $ Z( \pi / 4 ) $,
however, do not operate transversally even on the 7-qubit code.
Then, in order to implement the $ \pi / 8 $ gate
by a transversal measurement,
we exchange the $ Z ( \pm \pi / 4 ) $ rotation and the C-$ Z $ gates,
as shown in Fig. \ref{pi8-ope}.
As a result, the operation $ H Z( - \pi / 4 ) $
is equivalently implemented
by the preparation of $ Z( - \pi / 4 ) | +_{L} \rangle $
and the measurement with post-selection of $ I $ or $ S = Z( \pi / 2 ) $
operating transversally on the 7-qubit code.
In this way we can implement the $ H $, $ S $, $ \pi / 8 $ and C-$ Z $ gates
as a universal set by transversal measurements on logical clusters.

\begin{figure}
\centering
\scalebox{.3}{\includegraphics*[0cm,0cm][18cm,5cm]{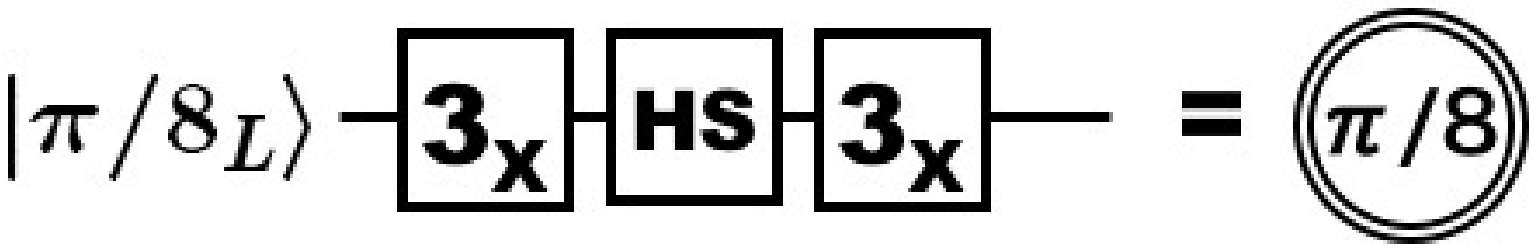}}
\caption{Fault-tolerant preparation
of $ Z( - \pi / 4 ) | +_{L} \rangle $.
The $ Z $ errors in the $ | {\pi / 8}_{L} \rangle $ are verified
by the first Box 3$ {}_X $, and the $ X $ errors,
which are converted to $ Z $ errors through the $ HS $ gate,
are detected by the second Box 3$ {}_X $.}
\label{pi8-pre}
\end{figure}

A fault-tolerant preparation of clean $ Z ( - \pi / 4 ) | +_{L}\rangle $
is shown in Fig. \ref{pi8-pre}.
This is based on the equivalence
\begin{equation}
Z( - \pi / 4 ) | +_{L} \rangle = e^{i \phi} HS | {\pi / 8}_{L} \rangle ,
\end{equation}
where $ \phi $ is a certain phase.
The logical $ | {\pi / 8}_{L} \rangle
= \cos ( \pi / 8 ) | 0_{L} \rangle + \sin ( \pi / 8 ) | 1_{L} \rangle $
is first encoded with physical qubits as given in Ref. \cite{Knill98},
where an ancilla cat state and physical C-$ H $ gates are used.
Then, it is verified against the $ Z $ errors by the first Box 3$ {}_X $,
and subsequently rotated by the $ HS $ gate.
The $ X $ errors in the $ Z( - \pi/4 ) | +_{L} \rangle $,
which are converted to $ Z $ errors through the $ HS $ gate,
are also detected by the second Box 3$ {}_X $.
As a result, we obtain the desired $ Z( - \pi/4 ) | +_{L} \rangle $
of as high fidelity as the level-2 verified $ | +_{L} \rangle $
in Fig. \ref{cluster-pre}.

\begin{figure}
\centering
\scalebox{.3}{\includegraphics*[0cm,0cm][25cm,20cm]{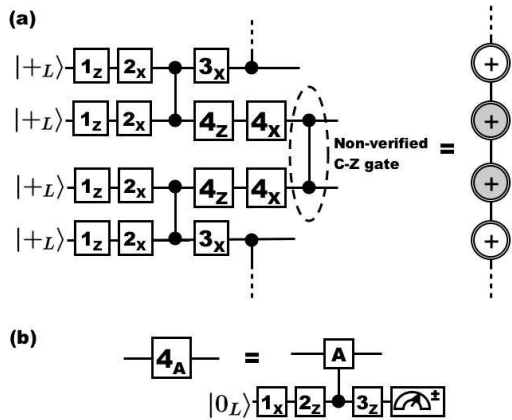}}
\caption{Connecting sub-clusters.
(a) The boundary qubits (shaded circles)
to be connected with non-verified C-$ Z $ gates
should be checked by Box 4's at a higher fidelity level.
Once the computation starts,
the cluster states cannot be verified by post-selection.
(b) Box 4 for the level-4 verification,
where the remaining $ X $ errors in the level-2 $ | 0_{L} \rangle $
are blocked by the Box 3 to propagate to the data qubit.
}
\label{connection}
\end{figure}

\begin{figure}
\centering
\scalebox{.3}{\includegraphics*[0cm,0cm][18cm,7.5cm]{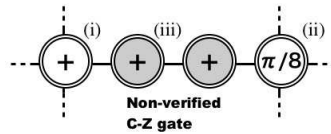}}
\caption{A logical cluster state for a simulation
to calculate the logical error probability
by performing a transversal measurement
on the qubits (i), (ii) and (iii), respectively.}
\label{cluster}
\end{figure}

As for the scalability of time direction,
memory errors should be considered.
If the entire cluster is prepared at the beginning,
the qubits are exposed to memory errors for a long time
until all the measurements are completed.
To overcome this difficulty we divide the entire cluster
of $ K \times Q $ matrix ($ Q $ operations for each of $ K $ qubits)
into consecutive $ K \times q $ sub-clusters or layers
\cite{OWC,Rausen03,Niel04,ND05}.
The computation steps $ q $ of each layer may be taken relevantly
so that the effect of memory errors does not exceed
those of gate and measurement errors.
The computation is performed as follows:
(1) The layer $ {\mathcal C}_1 $ is first prepared by the method (I).
(2) The logical qubits in $ {\mathcal C}_1 $
are measured transversely one after another.
In parallel the next layer $ {\mathcal C}_2 $ is prepared.
(3) Before the qubits of the last step in $ {\mathcal C}_1 $ are measured,
we connect them with C-$ Z $ gates
to the corresponding qubits in $ {\mathcal C}_2 $
(see Figs. \ref{connection} and \ref{cluster}).
These processes are repeated until the entire computation is completed.
Note here that the connections of the layers are not verified
while they are fault-tolerant with transversal C-$ Z $ gates.
This is because we are not able to verify the clusters by post-selection
once the computation starts by measuring them.
These non-verified C-$ Z $ gates, however, bring some error chances
to lower the noise threshold.
In order to improve this unavoidable situation
we apply the level-4 verification with Box 4$ {}_Z $ and Box 4$ {}_X $
to the qubits to be connected with non-verified C-$ Z $ gates,
as shown in Fig. \ref{connection}.
The Box 4 really has a higher fidelity,
since the Box 3 is inserted through the C-$ A $ gate
to check even the few errors which may escape from the level-2 verification
of $ | 0_{L} \rangle $.

This parallelism is profitable not only
for suppressing the effect of memory noise,
but also for the scalability of space direction
in the present scheme with preparation by post-selection.
We further divide each layer $ {\mathcal C}_l $
into smaller $ k \times q $ sub-clusters $ {\mathcal C}_l (n) $.
Then, non-verified C-$ Z $ gates,
as shown in Fig. \ref{connection}, are used
to connect these sub-clusters $ {\mathcal C}_l (n) $
to construct the layer $ {\mathcal C}_l $.
The level-4 verification is applied to the boundary qubits
for the connection.
This construction of layers is useful
to restrain the inflation of physical resource.
Let $ f $ be the mean number of physical qubits and gates
used to prepare one logical qubit through verification,
and $ p_{v} $ be the mean success probability
for the verification of one logical qubit having some branches
of C-$ Z $ connection.
Then, each $ k \times q $ sub-cluster can be prepared through verification
by making $ N / p_{v}^{kq} $ trials ($ N \sim 10 $) in parallel
with almost unit success probability ($ \geq 1 - e^{- N} $).
The $ K \times Q $ entire cluster is constructed
by connecting deterministically
$ (K/k)(Q/q) $ of these $ k \times q $ sub-clusters
with non-verified C-$ Z $ gates.
This construction requires the physical resource of qubits and gates
being roughly proportional to $ KQ $ as
\begin{equation}
{\cal R} \sim (K/k) (Q/q) (kqf) ( N / p_{v}^{kq} )
= N ( f / p_{v}^{kq} ) KQ .
\label{eqn:resource}
\end{equation}
Note that the resource would grow exponentially with $ p_{v} < 1 $
if $ k = K $ (and $ q = Q $) to prepare at once
the whole of each layer (the entire cluster).
The size $ kq $ of a sub-cluster may be chosen
by considering $ p_{v} ( p_{e} ) $ in an actual computation.
The minimum choice may be a level-3 qubit
having four branches to level-4 qubits
(a $ 3 \times 3 $ sub-cluster with the four corners deleted).

We have made a simulation for preparation and measurement
of a logical cluster state
to calculate the noise threshold for the present scheme.
The same noise parameterization is used as in Ref. \cite{Knill05b}.
Let $ p_{e} $ be the error probability per one physical operation.
A single-qubit operation has the probability $ 4p_{e}/5 $
for each of the 3 Pauli errors $ X $, $ Y $, $ Z $.
A two-qubit control gate has the probability $ p_{e}/15 $
for each of the 15 Pauli errors $ X \otimes I $, $ X \otimes Y $, and so on.
The errors in preparation and measurement of a physical qubit
are simulated by flipping the prepared state or the measurement result
with the probability $ 4p_{e}/15 $.
We have assumed for simplicity
that the probability of memory error $ p_{m} $ per one time step
is practically included into $ p_{e} $ of operation error.
This will be the case
if $ p_{m} $ is sufficiently smaller than $ p_{e} $,
satisfying the condition $ q , m < p_{e} / p_{m} $
for the width of a sub-cluster $ q $
and the number of waiting time steps $ m $
(3 or less in the present method)
between the gate operations in the preparation process (I).

\begin{figure}
\centering
\scalebox{.385}{\includegraphics*[4cm,2cm][12cm,14.5cm]{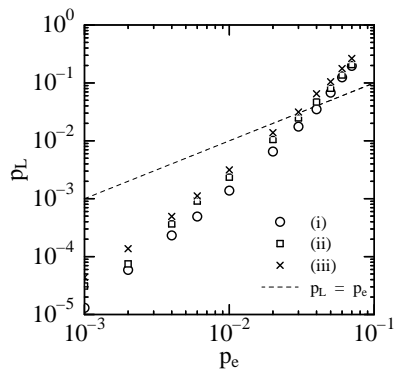}}
\caption{The logical error probability $ p_{L} $ is plotted
as a function of the physical error probability $ p_{e} $
for the cases (i), (ii) and (iii) as shown in Fig. \ref{cluster}.}
\label{threshold}
\end{figure}

The following operations are specifically simulated
by preparing a logical cluster state with verification
as shown in Fig. \ref{cluster}:
(i) A logical qubit $ | +_{L} \rangle $ having four branches
is measured transversally for the $ H $ gate operation.
(ii) Another logical qubit $ Z( - \pi / 4 )| +_{L} \rangle $
is measured transversally for the $ \pi / 8 $ gate operation.
(iii) Two level-4 verified qubits (shaded) are connected
with a non-verified C-$ Z $ gate,
and the left one is measured transversally
to teleport the quantum data fault-tolerantly between the sub-clusters.
Here we adopt the 7-qubit code with distance 3.
The logical error probability $ p_{L} $
that the measured qubit has two or more errors
after the transversal measurement is plotted
as a function of the physical error probability $ p_{e} $
in Fig. \ref{threshold}
for the cases (i) with circles, (ii) with boxes and (iii) with crosses,
respectively.
It is seen that the case (iii) of non-verified C-$ Z $ connection
actually determines the threshold, as expected.
These results really indicate
that fault-tolerant computation can be implemented
with a rather high noise threshold $ \approx 3 \% $.
For instance, if $ p_{e} \approx 1 \% $,
the error rate is reduced to $ p_{L} \approx 2 \times 10^{-3} $
by using the present method.
Then, the usual circuit-based methods
may be used on the upper levels of the QEC code concatenation.
As for the physical resource, we estimate $ f \approx 500 $
and $ p_{v} \approx 0.7 $ for $ p_{e} \approx 1 \% $.
Then, a computation of a size $ K \times Q = 100 \times 1000 $
will be implemented with about $ 10^{12} $ ($ 10^{10} $)
physical qubits and gates
by using $ k \times q = 6 \times 4 $ ($ 3 \times 3 $) sub-clusters,
as given in Eq. (\ref{eqn:resource}) with $ N \sim 10 $.

The above threshold result is about an order of magnitude higher
than those by the usual circuit-based QEC methods.
This is because in the present scheme
the recovery operation is not required
owing to the unique way of quantum processing in the cluster model.
The syndrome detection and error correction
by themselves include many operations to cause errors.
In contrast, once logical clusters of high fidelity are prepared
by post-selection through verification,
only transversal measurements are made on the clusters.
This reduces error chances significantly to achieve a high noise threshold.

In summary, we have investigated
a scheme of fault-tolerant quantum computation
based on the cluster model.
Logical qubits are encoded by a suitable code
such as the Steane's 7-qubit code.
Cluster states of logical qubits are prepared by post-selection
through verification at high fidelity level,
where the unsuccessful ones are discarded without recovery operation.
Then, gate operations are implemented
by transversal measurements on the prepared logical cluster states.
The noise threshold is improved significantly
by making the high fidelity preparation and transversal measurement.
It is estimated to be about $ 3 \% $ by a numerical simulation,
which is quite higher than those of the circuit-based QEC methods.

This work is supported by International Communications Foundation (ICF).

\end{document}